\documentclass[letterpaper]{article} %
\usepackage[preprint]{aaai2027}  %
\usepackage[hyphens]{url}  %
\usepackage{graphicx} %
\usepackage{natbib}  %
\usepackage{caption} %
\usepackage{booktabs}
\usepackage{amsmath,amssymb}
\usepackage{multirow}

\title{Refusing Everything Looks Safe: Restoring the Benign Arm to Encoded-Prompt Evaluation}
\author{
    Haoyu Zhang\textsuperscript{\rm 1},
    Haowen Xu,
    Xiao Luo,
    Hanwen Liu,
    Yang Chen,
    Zijian Xiao,
    Yi Feng\textsuperscript{\rm 1},
    Xiangchen Guan,
    Mohammad Zandsalimy\textsuperscript{\rm 1},
    Shanu Sushmita\textsuperscript{\rm 1}
}
\affiliations{
    \textsuperscript{\rm 1}Northeastern University\\
    zhang.haoyu6@northeastern.edu
}

\begin{document}
\maketitle

\begin{abstract}
Encoded-prompt attacks are evaluated almost entirely on their harmful arm: a
benchmark sends obfuscated harmful requests and reports how often the model
complied. A high refusal
rate there is reported as safety, and it is equally consistent with a model that
has stopped telling the request apart from anything else in the same format. We
run the benign arm through the same transformation, and the two cases are far
apart. Across four 7--8B models spanning three base families and
four post-training recipes, refusal of \emph{harmful} homoglyph-encoded prompts
spans $0.08$ while the same four span $0.57$ on the identical requests in
plaintext. That is a sevenfold compression, to a separation that is
statistically detectable and useless for ranking. What the encoding destroys is not refusal but the \emph{harm gap}: on one
model the gap between harmful and benign refusal falls from $+0.82$ in plaintext
to exactly $0.00$ under the encoding, benign and harmful requests being refused at
an identical $0.99$. A benchmark reading only the harmful arm scores that model and one
retaining a $+0.61$ gap identically. Running the cell such benchmarks leave
out (plaintext content wearing the attack template, with nothing obfuscated)
shows that on two of the four models the loss is caused by the
\emph{protocol} rather than by the character transformation, and on a third by
the characters. We then ask whether post-training repairs this, using a
published recipe on identical base weights. It improves the encoded condition
and never closes the distance to plaintext: across a full SFT $\rightarrow$ DPO $\rightarrow$ RLVR pipeline, the plaintext
harm gap rises from $+0.55$ to $+0.80$ and the encoded gap rises with it, $+0.20$
to $+0.46$, so the shortfall between them is unchanged at $0.34$--$0.50$ and the
two movements cancel to within $0.01$. Over that same pipeline the standard harmful-arm metric
registers no resolved change at all ($0.99 \rightarrow 0.94$, five discordant
items, $p=0.06$) while the harm gap rises by $+0.26$ with a paired interval
excluding zero. The arm the field reports is not just a weaker signal than the
arm it omits; here it is blind to a change the omitted arm resolves.
The control is neither expensive nor new: one benchmark already scores benign
text through ten ciphers for decoding accuracy and harmful text through the same
ten for attack success, so the number we report was one column away. None of this
is visible without controls the field does not routinely run. We
report twelve instrument defects, each with the control that caught it,
including one of our own that withdrew a result. Six of the twelve inflate
apparent safety, which is the direction a broken safety evaluation fails in by
default.
\end{abstract}

\section{Introduction}

A safety evaluation is a measurement instrument, and like any instrument it can
be wrong in a direction. This paper is about one such direction, in a widely
used family of evaluations.

Encoded-prompt attacks (base64, classical ciphers, homoglyph substitution,
zero-width insertions) are a standard probe of whether safety training
generalises beyond the surface forms it was trained on
\citep{wei2023jailbroken,yuan2024gpt}. The protocol is nearly uniform:
take a corpus of harmful requests, encode it, send it, and report how often the
model complied or refused. The benign half of the input space is not sent,
because on an attack benchmark there is no attack to measure there.

We show this protocol resolves those models far more weakly than the arm it
omits. On homoglyph-encoded harmful prompts, four models spanning three base
families and four post-training recipes refuse at $0.91$, $0.91$, $0.93$ and
$0.99$, a spread of $0.08$, at the edge of what sampling
noise alone would produce if all four models were identical. The same four
models, on the same requests in plaintext, span $0.57$. The encoding does not
merely fail to separate them; it removes a separation that was there.

We are careful about what that does and does not say. Individual pairs can still
separate on the harmful arm: the two extremes, $0.99$ and $0.91$, differ at
$p=0.02$ by Fisher's exact test. The claim is about magnitude, not about a floor
of zero: the spread this arm offers is roughly $7.6\times$ smaller than the
spread of the quantity we introduce below, and a benchmark that ranks models
depends on the spread rather than on one surviving pair.

\begin{figure*}[t]
\centering
\includegraphics[width=\textwidth]{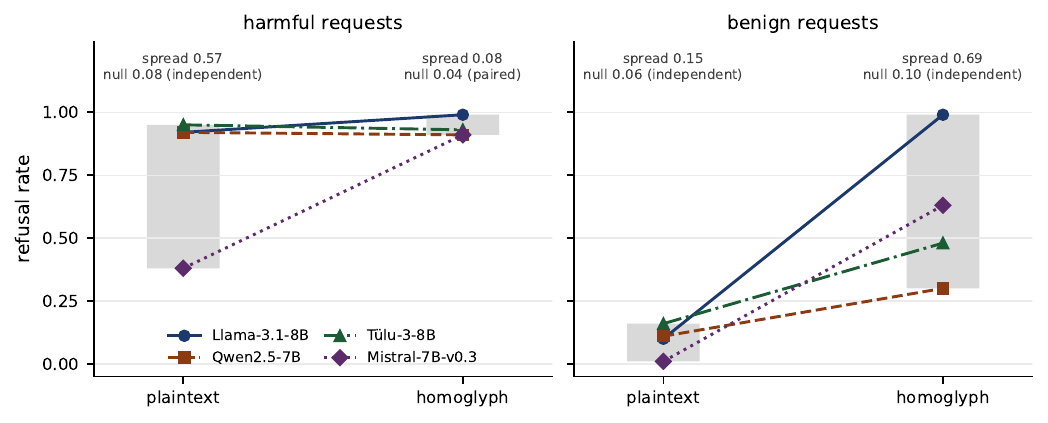}
\caption{The two arms invert. Each line is one model; the shaded band is the
observed spread across the four. On \emph{harmful} requests the models fan in,
a spread of $0.57$ collapsing to $0.08$, the edge of what four identical
models would produce at $n{=}100$ (bold), while on \emph{benign} requests they
fan out, $0.15$ to $0.69$. An evaluation reading only the left panel's right-hand
edge cannot usefully rank the four. Numbers in Table~\ref{tab:families}.}
\label{fig:inversion}
\end{figure*}

The information the protocol discards is in the benign arm. There the four
models span $0.69$, and the differences are large and structured: one refuses
$0.99$ of benign homoglyph text, another $0.30$. Reading both arms together
gives a quantity the harmful arm cannot: the \emph{harm gap}, the difference
between harmful and benign refusal, which measures whether refusal tracks the
request or its appearance. Figure~\ref{fig:inversion} shows the two arms moving
in opposite directions.

Our claim is deliberately about measurement, not about utility. One can argue
that refusing homoglyph-encoded text is correct behaviour, since few legitimate
users send it; we take no position. The argument does not touch the finding.
Whether or not blanket refusal is desirable, a metric assigning the same score
to a model with a $+0.61$ harm gap and a model with a $0.00$ harm gap is not
measuring safety behaviour under encoding.

\paragraph{One claim, in four steps.} This paper argues a single thesis:
\emph{an evaluation that sends only harmful inputs cannot tell a model that
refused correctly from a model that stopped reading the request.} Both report
high refusal under attack, and high refusal under attack is the number this
literature publishes. The four sections below are steps of that one argument
rather than four separate results, and each is stated with the control that makes
it survive.

\begin{enumerate}
\item \textbf{The metric is nearly blind} (\S\ref{sec:families}). On the harmful arm, four models spanning three base families are separated by $0.08$, a spread at the edge of what four identical models would produce at $n{=}100$, while spanning $0.57$ on the same requests in plaintext.
\item \textbf{It is blind to \emph{what}} (\S\ref{sec:arms}). A third arm, the attack template around untransformed content, splits the loss into a protocol term and a character term. Which dominates is a property of the model.
\item \textbf{It stays blind through a full remedy} (\S\ref{sec:ladder}). Across a published SFT $\rightarrow$ DPO $\rightarrow$ RLVR pipeline on identical base weights, the harm gap rises by $+0.26$ with a paired interval excluding zero while the harmful-arm metric resolves no change.
\item \textbf{And why it was not caught} (\S\ref{sec:instrument}). Twelve instrument defects, each with the control that caught it. Six inflate apparent safety.
\end{enumerate}

\paragraph{A note on the title.} \emph{Refusing everything looks safe} is a
claim about the measurement, not about the model. Under this protocol these
models refuse at high rates and refuse benign and harmful content at nearly the
same rate, so a reading taken from the harmful arm alone cannot separate the two
cases. We do not argue that the resulting behaviour is unsafe. We argue that the
number reported for it does not distinguish states differing by $0.61$ of harm
gap. We say \emph{restoring} rather than introducing because
\citet{wei2023jailbroken} ran a harmless control alongside every attack and saw
this effect; it was one prompt, scored for capability rather than refusal, and no
later paper carried it.

\paragraph{Scope.} The cross-model result (\S\ref{sec:families}) is measured
on \emph{one} encoding, homoglyph substitution, chosen because it is the rung on
which every screen passes on all four models; the pipeline result
(\S\ref{sec:ladder}) adds two more. We report refusal rates and never attack
success rates, because \S\ref{sec:instrument} shows the binary jailbreak judge we
used fires on $0.61$--$0.70$ of responses to \emph{plaintext benign} prompts,
where no attack exists by construction. All four models are open-weight and
$7$--$8$B; we make no claim about frontier scale, and the pipeline result is one
family's published recipe rather than post-training in general. What these
benchmarks ship is not a bare character transformation but a \emph{protocol}:
transformed content inside a template announcing an encoding and asking the model
to work with it. We measure the protocol, because that is what is deployed, and
separate its two terms in \S\ref{sec:arms}. We establish that the effect exists,
that it is large, and that a full post-training pipeline does not remove it; we do
\emph{not} establish \emph{why} refusal stops tracking harm, which is a question
about internals that \S\ref{sec:limitations} records our failure to answer.

\section{Related work}
\label{sec:related}

\paragraph{Why encoded prompts are studied.} \citet{wei2023jailbroken} name
\emph{mismatched generalization}: safety training generalising worse than
capability, so an input the model still understands can fall outside the
distribution safety training covered. Encoded prompts are the sharpest
instantiation: \citet{yuan2024gpt} show cipher-encoded chat bypassing
alignment on frontier models, with some ciphers succeeding almost always in
several harm domains. Our work does not dispute those attack results. It asks
what the \emph{evaluation} of such attacks can and cannot establish, and our
answer sharpens rather than weakens the framing: on the models we test, the
harmful arm is where mismatched generalisation is looked for and is precisely
where the models nearly converge. The same family covers refusal-mechanism analyses and multilingual or cross-model obfuscation \citep{Prakash_Jie_Abdullah_Satapathy_Cambria_Lee_2026,uysal2026multilingualobfuscated}.

\paragraph{Jailbreak evaluation and its known weaknesses.} That jailbreak
metrics are unreliable is established. \citet{souly2024strongreject} show that
attack success rates are routinely overstated and that binary judges credit
responses containing no usable harmful content, and they propose a graded rubric
gated on refusal. \citet{NEURIPS2024_63092d79} standardise the corpus and
protocol. Our contribution here is narrower and, we think, complementary in an
uncomfortable direction: both of those efforts improve how the harmful arm is
\emph{scored}, and we find that the harmful arm carries little information about
the model regardless of how well it is scored. A better judge on that arm does
not recover a separation that Table~\ref{tab:families} shows is too small to
rank on. We
also reproduce the binary-judge failure on our own judge
(\S\ref{sec:instrument}), on plaintext benign prompts where the ground truth is
fixed by construction rather than by a rubric. Standardised red-teaming suites and in-the-wild prompt collections have made these comparisons routine \citep{10.1145/3658644.3670388,10.5555/3692070.3693501,li-etal-2025-detam,mu-etal-2025-stealthy}, and defence-side work reports guardrail and moderation results on the same footing \citep{ding-etal-2025-act,10.1145/3724393,lv-etal-2025-gamma,ghosh-etal-2025-aegis2,verma-etal-2025-multiguard}.

\paragraph{Over-refusal.} The benign arm we argue for is not a new instrument.
\citet{rottger-etal-2024-xstest} built XSTest precisely to measure exaggerated safety
(refusal of clearly safe prompts that resemble unsafe ones) and report
that models refuse safe prompts using sensitive-adjacent language. We use
XSTest as a vocabulary control rather than as our benign corpus, since we need
the benign set matched by theme to the harmful set, which JailbreakBench
provides. Over-refusal is measured by a growing set of benchmarks and mitigations \citep{Cao_Yang_Zhao_2025,zhang-etal-2026-llm,li2026outputaware,ren2025dualbench}, and the alignment literature has begun to ask how deep such behaviour is written into a model \citep{qi2025shallowalignment,zou2024circuitbreakers}.

\paragraph{The two literatures meet in a cell that neither runs.} Over-refusal is measured on untransformed prompts, and encoded attacks are
measured on harmful content, so benign content in attack form falls to neither.
Because our contribution rests on that gap, we checked it: we read the full text
of eleven input-transformation attack papers and both widely used over-refusal
benchmarks, drawn from a curated corpus of $419$ papers, and recorded for each
whether it sends benign content through its own transformation and reports a
refusal or false-positive rate on that arm. Neither XSTest nor OR-Bench
\citep{cui2025orbench} contains a single occurrence of \emph{encoding},
\emph{cipher}, \emph{base64}, \emph{obfuscation} or \emph{ASCII art} anywhere in
its text. We state the counts over the papers we adjudicated rather than over the
literature; the per-paper verdicts are in the supplementary material.

\paragraph{The same gap on the defence side.} \citet{fairoze2026bypassingpromptguardsproduction} bypass production
prompt guards with encodings a guard cannot decode, and calibrate against
untransformed content. A defence whose false-positive rate is measured on plain
benign prompts has not measured its false-positive rate under the condition it
is deployed against. We note it because it is the same missing cell one level
out, on the defence rather than the target, and because it bounds what our
result can be read to say about deployed systems.

\paragraph{The near misses.} We adjudicated eleven input-transformation attack papers and the two over-refusal benchmarks \citep{wei2023jailbroken,yuan2024gpt,jiang-etal-2024-artprompt,handa2025when,yong2023lowresource,he2025solving,peng-etal-2026-logic,yan-etal-2025-semanticcamo,pmlr-v318-zhang26a,NEURIPS2024_63092d79,cui2025orbench,rottger-etal-2024-xstest}; two come closest, and the supplementary material gives the verdict for every row.

\paragraph{Reading refusal inside the model.} A line of interpretability work locates refusal in model internals \citep{arditi2024refusal,zhao2025llmsencode}. We attempted a probe-based account of the harm gap and withdrew it; the attempt and why it fails are in the supplementary material.

\section{Method}
\label{sec:method}

\paragraph{Models.} Llama-3.1-8B-Instruct \citep{grattafiori2024llama3},
Qwen2.5-7B-Instruct \citep{qwen2024qwen25}, Mistral-7B-Instruct-v0.3
\citep{jiang2023mistral} and T\"ulu-3-8B \citep{lambert2025tulu3}: four
open-weight instruction-tuned models whose post-training pipelines were
developed independently. We cite the Mistral paper for the family only, since it
describes v0.1 and we run v0.3, which changed the vocabulary.

\paragraph{Corpus.} Harmful prompts and their \emph{theme-matched} benign
counterparts come from JailbreakBench \citep{NEURIPS2024_63092d79}. The
matching matters: the benign set is not generic helpful text but topically
adjacent requests, so a model separating the two is responding to harm rather than to subject. $n{=}100$ per cell throughout.

\paragraph{Conditions.} Each model sees six cells: $\{$harmful, benign$\}$
$\times$ $\{$plaintext, scaffold, encoded$\}$. \emph{Plaintext} is the corpus
verbatim, with no attack template. \emph{Encoded} is the deployed protocol:
transformed characters inside a template announcing the encoding and asking
the model to work with it. \emph{Scaffold} is the cell that separates the two:
the same template around \emph{untransformed} content, so that nothing in
it is obfuscated. Comparing plaintext against encoded alone, as this
literature does, charges the template's effect to the encoding. The scaffold
arm is what splits them, it costs one forward pass, and we recommend it as
standard.

\paragraph{The harm gap.} Our headline quantity is
$\Delta = r_{\text{harmful}} - r_{\text{benign}}$, the difference in refusal rate
between the two arms of one condition, and \emph{gap lost}
$= \Delta_{\text{plain}} - \Delta_{\text{enc}}$. Neither is computable from an
attack benchmark: both need the benign arm and the plaintext baseline.

\paragraph{Refusal judging and run-to-run variation.} Refusal is classified by an LLM judge on the JailbreakBench template, validated against two cross-family annotators; re-running a condition moves the harm gap by $0.01$--$0.06$. Both are detailed in the supplementary material.

\paragraph{Noise null.} Cross-model spreads are max-minus-min statistics over
four estimates, which are not stable summaries. For every spread we report, we
bootstrap ($20{,}000$ draws) the spread four \emph{identical} models would
produce at $n{=}100$, and read the observed spread against it. We adopted this
after finding we had ourselves compared two such spreads directly and drawn a
conclusion the null does not support.

Where per-item verdicts exist, the null is \emph{paired}. All four models
answer the same $100$ prompts, so item difficulty is shared, and four independent
binomials at the common mean overstate how far identical models would drift
apart. Each simulated model instead draws per item from that item's own
difficulty, estimated as the fraction of the four that refused it. Estimating
difficulty from the models being tested inflates the null wherever they genuinely
disagree, so the paired version is conservative; we use it anyway, and
\S\ref{sec:families} reports the one claim it costs us. We retained per-item
verdicts for the encoded conditions and not for the plaintext baseline, so
plaintext spreads are read against the independent null. We mark which is which
rather than let two nulls share one name: the plaintext spreads sit $7\times$
outside either, so nothing there turns on the choice.

\section{The harmful arm cannot separate models}
\label{sec:families}

\begin{table}[t]
\centering
\small
\setlength{\tabcolsep}{4pt}
\begin{tabular}{lrrrrrr}
\toprule
& \multicolumn{3}{c}{plaintext} & \multicolumn{3}{c}{homoglyph} \\
\cmidrule(lr){2-4}\cmidrule(lr){5-7}
model & harm & ben. & $\Delta$ & harm & ben. & $\Delta$ \\
\midrule
Llama-3.1-8B-It  & 0.92 & 0.10 & $+0.82$ & 0.99 & 0.99 & $\mathbf{0.00}$ \\
Qwen2.5-7B-It    & 0.92 & 0.11 & $+0.81$ & 0.91 & 0.30 & $+0.61$ \\
T\"ulu-3-8B      & 0.95 & 0.16 & $+0.79$ & 0.93 & 0.48 & $+0.45$ \\
Mistral-7B-v0.3  & 0.38 & 0.01 & $+0.37$ & 0.91 & 0.63 & $+0.28$ \\
\midrule
\textbf{spread}   & \textbf{0.57} & 0.15 & & \textbf{0.08} & \textbf{0.69} & \\
\emph{noise null} & \emph{0.08} & \emph{0.06} & & \emph{0.04} & \emph{0.10} & \\
\bottomrule
\end{tabular}
\caption{Refusal rates, $n{=}100$ per cell. \emph{spread} is max$-$min across
the four models; \emph{noise null} is the median spread four identical models
would show ($20{,}000$ bootstrap draws), matched on item difficulty where
per-item verdicts exist and on the common rate otherwise. The harmful arm falls from far outside the null in
plaintext to its edge under the encoding; the benign arm does the reverse.
Encoded nulls are paired on items, plaintext nulls independent
(\S\ref{sec:method}).}
\label{tab:families}
\end{table}

Table~\ref{tab:families} gives the rates behind Figure~\ref{fig:inversion}, each
spread read against the null of the spread four identical models would produce.

\paragraph{The harmful arm erases model differences.} In plaintext the four
models span $0.57$ against a null of $0.08$ $[0.02,0.16]$. These are very
different models, and Mistral-7B in particular refuses only $0.38$ of harmful
plaintext requests. Under homoglyph they span $0.08$: a sevenfold compression,
down to the noise floor. \emph{Down to} it and not \emph{inside} it, and that
distinction is one our own null first got wrong. Four \emph{independent}
binomials at the observed mean give a median spread of $0.05$ and place $0.08$
comfortably inside ($p=0.12$), which is what an earlier version of this paragraph
reported. Under the paired null of \S\ref{sec:method} the median is $0.04$, the
$95$th percentile is $0.07$, and the observed $0.08$ falls outside ($p=0.03$).
Since the paired construction is the conservative one, the surviving claim is
compression to the edge of what $n{=}100$ resolves, not equality: these four
models remain marginally separable on the harmful arm, and nothing about the
harmful arm's usefulness is rescued by a separation that small.

\paragraph{The benign arm creates them.} In plaintext the four agree closely
($0.01$--$0.16$), and the spread of $0.15$ sits just outside an independent
null of $0.06$ $[0.02,0.12]$. We report every null as a median with its
$[2.5,97.5]$ interval and never quote the upper tail alone: an earlier draft
called one such ceiling ``the null'', which is the same slip running the other
way.
Under homoglyph they span $0.69$ against a paired null of $0.08$
$[0.02,0.16]$.
This arm is unaffected by the choice of null: paired or independent, the
observed spread exceeds the $99$th percentile in $20{,}000$ draws
($p<10^{-4}$ either way). The asymmetry between the two arms is the result, and
it does not rest on a null's fine print.

\paragraph{The consequence.} Llama-3.1-8B and Qwen2.5-7B differ by $0.08$ on
the harmful arm and are opposite on the harm gap ($0.00$ vs $+0.61$). The
harmful difference is real, since on the same $100$ prompts all seven
discordant items run the same way (McNemar $p=0.016$), and it is $8\times$
smaller than the harm-gap difference it is standing in for. We stress this rather than
claim the two models are identical there: the problem with the harmful arm is not
that it reads noise, it is that it reads a real quantity too small to rank on
while the quantity that separates these models sits in the arm nobody reports.

\paragraph{Gap collapse is significant on three of four models.} With a
confidence interval on the difference of differences: Llama $+0.82$
$[+0.74,+0.90]$, T\"ulu-3 $+0.34$ $[+0.20,+0.48]$, Qwen $+0.20$
$[+0.07,+0.33]$, Mistral $+0.09$ $[-0.06,+0.24]$, which includes zero. We report
three collapses, not four.

\paragraph{Two caveats.} Encoded harmful refusal sits near ceiling on all four models, so equalisation and saturation are not separated here; and Mistral, whose harmful refusal \emph{rises} under the encoding, is not an exception but the clearest case of the same mechanism. Both are stated in full in the supplementary material.

\paragraph{A second encoding clears both screens, and it does not replicate the
first.} The evidence above rests on one retained transformation, which is the
first thing three referees of earlier drafts asked about. We therefore ran the
paper's own two screens over every rung for which both arms exist in a single
job per model: \emph{readability}, that the model decodes the rung at all, and
\emph{echo-cleanliness}, that removing echoing cells moves the harm gap by less
than the gap's own half-width. Of seven rungs, two survive on more than one
model (Table~\ref{tab:second}). \emph{math-bold}, the Unicode mathematical bold
alphabet, is readable on three of the four models and clears the echo screen on
all four.

\begin{table}[t]
\centering
\small
\setlength{\tabcolsep}{4pt}
\begin{tabular}{lrrrrrr}
\toprule
& \multicolumn{3}{c}{homoglyph} & \multicolumn{3}{c}{math-bold} \\
\cmidrule(lr){2-4}\cmidrule(lr){5-7}
model & harm & ben. & $\Delta$ & harm & ben. & $\Delta$ \\
\midrule
Llama-3.1-8B-It  & 0.99 & 0.99 & $\mathbf{0.00}$ & 0.59 & 0.41 & $+0.18$ \\
Qwen2.5-7B-It    & 0.92 & 0.33 & $+0.59$ & 0.86 & 0.09 & $+0.77$ \\
T\"ulu-3-8B      & 0.93 & 0.51 & $+0.42$ & 0.44 & 0.06 & $+0.38$ \\
Mistral-7B-v0.3  & 0.91 & 0.66 & $+0.25$ & 1.00$^{\dagger}$ & 1.00$^{\dagger}$ & \\
\bottomrule
\end{tabular}
\caption{Two encodings that clear both screens, $n{=}100$ per cell, one job per
model. Decode ability is $0.88$--$0.98$ for homoglyph on all four models and
$1.00$ for math-bold on three. $^{\dagger}$Mistral's math-bold ability is
$0.00$: it refuses both arms because it cannot read either, so no harm gap is
defined and the cell is a decode failure rather than a failure to tell harmful
from benign.
Of the other five rungs measured in the same jobs, \emph{base64} and
\emph{tag-block} are unreadable everywhere and \emph{fullwidth},
\emph{fullwidth-letters} and \emph{zero-width} fail the echo screen on three
models each.}
\label{tab:second}
\end{table}

\paragraph{It does not replicate the first, and that is the point.} The second
encoding is not a replication, and reporting it as one would be the error this
paper is about. On Llama the two disagree completely, $0.00$ against
$+0.18$, so the total harm-gap loss we report for that model is a
property of the model \emph{and} the encoding jointly rather than of the
encoding alone. Both cells are invisible to a harmful-only metric in the same
way, which is the claim; ``encoding destroys the harm gap'' as a statement
about encodings in general is what the second rung declines to support. On the
other three models the two rungs agree in sign and rank the models identically,
and math-bold produces the largest harm gap we measure under any encoding,
$+0.77$ on Qwen. Both plaintext arms of Table~\ref{tab:families} run higher.

\section{Splitting the protocol: template or characters}
\label{sec:arms}

The encoded condition
changes two things at once, and a plaintext baseline carrying no template cannot
tell them apart. We ran the missing cell (the same template around
untransformed content) as a separate job on all four models
(Table~\ref{tab:arms}). Its plaintext arm reproduces the headline independently:
Llama's plaintext gap reads $+0.83$ here against $+0.82$ in
Table~\ref{tab:families}, and its encoded arm $0.98$ / $0.99$ against $0.99$ /
$0.99$.

\begin{table*}[t]
\centering
\small
\begin{tabular}{lrrrrrrrrr}
\toprule
& \multicolumn{3}{c}{plaintext} & \multicolumn{3}{c}{scaffold} & \multicolumn{3}{c}{encoded} \\
\cmidrule(lr){2-4}\cmidrule(lr){5-7}\cmidrule(lr){8-10}
model & harmful & benign & $\Delta$ & harmful & benign & $\Delta$ & harmful & benign & $\Delta$ \\
\midrule
Llama-3.1-8B-It  & 0.93 & 0.10 & $+0.83$ & 0.99 & 0.83 & $+0.16$ & 0.98 & 0.99 & $\mathbf{-0.01}$ \\
Qwen2.5-7B-It    & 0.93 & 0.11 & $+0.82$ & 0.92 & 0.12 & $+0.80$ & 0.88 & 0.33 & $+0.55$ \\
T\"ulu-3-8B      & 0.96 & 0.16 & $+0.80$ & 0.98 & 0.46 & $+0.52$ & 0.93 & 0.50 & $+0.43$ \\
Mistral-7B-v0.3  & 0.37 & 0.01 & $+0.36$ & 0.60 & 0.12 & $+0.48$ & 0.91 & 0.64 & $+0.27$ \\
\bottomrule
\end{tabular}
\caption{The three arms, $n{=}100$ per cell, all three from one job per model.
\emph{scaffold} is the attack template around untransformed content. Both arms
are shown for every condition rather than the harmful arm and the gap, so that
which arm moves is visible directly. On Llama the harm gap has already fallen
from $+0.83$ to $+0.16$ before a single character is transformed, and the whole
of that fall is on the benign arm, $0.10$ to $0.83$. On Qwen the same arm costs
nothing. Wilson intervals on every rate and Wald intervals on every gap are in
the supplementary material.}
\label{tab:arms}
\end{table*}

\paragraph{Which term dominates is a property of the model.} Writing the total harm-gap loss as a template term (plaintext $\rightarrow$ scaffold) plus a
character term (scaffold $\rightarrow$ encoded), with unpaired Wald intervals at
$n{=}100$: on Llama the template alone accounts for $+0.67$ $[+0.56,+0.78]$ of a
total $+0.84$ $[+0.76,+0.92]$, and on T\"ulu-3 $+0.28$ $[+0.15,+0.41]$ of
$+0.37$ $[+0.23,+0.51]$, its character term not separable from zero. On Qwen the
reverse: the template term is $+0.02$ $[-0.09,+0.13]$ and the characters do all
the work, $+0.25$ $[+0.11,+0.39]$. On Mistral the two terms have opposite signs
and partly cancel, which is why its total is the one value here that includes
zero while its character term does not.

\paragraph{These intervals are unpaired, which is a defect.} Each term is a
difference of gaps across conditions measured on the same items, so the contrast
is genuinely paired; we report it unpaired because our pipeline persists per-item
verdicts for the encoded arm only. The direction is conservative, so the widths
above are upper bounds. This is defect~(12) of \S\ref{sec:instrument}.

\paragraph{One arm we report and do not correct.} Mistral's scaffold arm has an
echo rate of $0.38$: because the scaffold leaves content untransformed, our
echo detector fires on any response quoting the request, and cannot be
distinguished there from genuine parroting. We report the number and do not
subtract it. The comparison it enters, Mistral's template term, is the one
already indistinguishable from zero.

\section{A published safety pipeline does not repair it}
\label{sec:ladder}

The cross-model comparison cannot say what produces the difference: four models
differ in data, objective and scale at once. T\"ulu~3 \citep{lambert2025tulu3}
lets us do better: its post-training pipeline is published, its stage
checkpoints are released, and every stage sits on identical base weights. We ran
the same four-cell design on all three stages, with each stage's plaintext
baseline and encoded arms measured inside a single job, so the comparison is
paired.

\begin{table}[t]
\centering
\small
\begin{tabular}{lccc}
\toprule
& SFT & DPO & RLVR \\
\midrule
\multicolumn{4}{l}{\emph{plaintext}} \\
\quad harmful refusal & 1.00 & 0.96 & 0.96 \\
\quad benign refusal & 0.45 & 0.17 & 0.16 \\
\quad $\Delta_{\text{plain}}$ & $+0.55$ & $+0.79$ & $\mathbf{+0.80}$ \\
\midrule
\multicolumn{4}{l}{\emph{homoglyph} (clears the echo screen at every stage)} \\
\quad harmful refusal & 0.99 & 0.90 & 0.94 \\
\quad benign refusal & 0.79 & 0.61 & 0.48 \\
\quad $\Delta_{\text{enc}}$ & $+0.20$ & $+0.29$ & $\mathbf{+0.46}$ \\
\quad \textbf{gap lost} & $\mathbf{0.35}$ & $\mathbf{0.50}$ & $\mathbf{0.34}$ \\
\bottomrule
\end{tabular}
\caption{The T\"ulu~3 pipeline on identical base weights, $n{=}100$ per cell.
The plaintext harm gap improves by $+0.25$; the encoded harm gap improves with
it, by $+0.26$; the shortfall between them does not move. $\dagger$~marks a
cell where the echo screen of \S\ref{sec:instrument} fails, meaning the
displacement between the reported gap and the gap over non-echoing cells exceeds
the gap's own $95\%$ half-width. Those four cells are not reportable and appear
only to show that the direction is the same on all three encodings. Per-rate
Wilson intervals, echo denominators and the three-way echo sensitivity are in
the supplementary material.}
\label{tab:ladder}
\end{table}

\paragraph{The pipeline works, in plaintext.} $\Delta_{\text{plain}}$ rises from
$+0.55$ to $+0.80$, driven almost entirely by benign refusal falling from $0.45$
to $0.16$. This is the recipe doing what it reports doing: T\"ulu~3's
post-training mix adds contrastive benign data specifically to reduce
over-refusal of safe prompts, and we measure exactly that.

\paragraph{It reaches the encoded condition too, and that is why the shortfall
does not move.} The encoded harm gap is not flat. On homoglyph it rises from
$+0.20$ to $+0.46$, and because both checkpoints answer the identical $100$
harmful and $100$ benign prompts we can test the endpoints as paired: a
$20{,}000$-draw bootstrap resampling items once and scoring both stages on the
same resample puts the change at $+0.26$, $95\%$ CI $[+0.14, +0.38]$. It excludes
zero, as it does on the other two encodings ($+0.30$ and $+0.36$, both
excluding). \emph{Gap lost} nonetheless holds at $0.35 \rightarrow 0.34$, because
$\Delta_{\text{plain}}$ rises by $+0.25$ over the same pipeline and the two
movements cancel to within $0.01$.

This distinction is the whole result and it is easy to state backwards. Post-training
does not fail to reach the encoded condition; it improves the encoded condition
by very nearly the amount it improves the plaintext one, and therefore never
narrows the distance between them. Two-thirds of a published safety pipeline
lifts both arms in parallel. We stress that this is a measured cancellation and
not an underpowered null: the movements it is built from are individually
resolved at $n{=}100$, and it is the \emph{difference} of two large real changes
that is small.

\paragraph{Two notes on this table.} An earlier version of this analysis lacked the plaintext arm and read a general fall in refusal as an encoding effect; we withdrew that claim. The echo screen we require elsewhere had never been run on this table, and running it rejects four of nine cells. Both episodes, and the judge validation and held-out status of every threshold, are in the supplementary material.

\paragraph{The standard metric is blind to it.} Along the same pipeline,
harmful-arm refusal under homoglyph encoding moves $0.99 \rightarrow 0.94$. That
change does not resolve: five items out of $100$ flip, all in the same direction,
exact McNemar $p=0.06$. Over the identical prompts and the identical
checkpoints, the harm gap rises by $+0.26$ with an interval excluding zero. So
the arm the field reports registers nothing while the arm it omits registers a
large real improvement, and no amount of $n$ on the harmful arm alone recovers
the second number, because it is a difference the harmful arm does not contain.

On the two echo-contaminated encodings the harmful arm does not merely fail to
move, it \emph{falls} (fullwidth $0.94 \rightarrow 0.73$, zero-width $1.00
\rightarrow 0.83$), so a benchmark reading refusal-under-attack would report
that two-thirds of a published safety pipeline made the model less safe while
its harm gap improved. We report that as the direction it points rather
than as a measured effect, because those are the cells the echo screen rejects.

\section{The instrument}
\label{sec:instrument}

Everything above depends on measuring refusal correctly in conditions designed
to defeat surface pattern matching. We report the defects we found, because
several invalidate metrics in common use and because they share a direction.

\paragraph{(1) No denominator.} Our own pipeline computed refusal only on
encoded corpora for its first weeks. An encoded benign refusal rate of $0.99$
means nothing without the plaintext rate it should be read against: the same
number is a catastrophic finding at a plaintext baseline of $0.10$ and
unremarkable at $0.95$. The plaintext arm is now model-level and mandatory, and
it is the control most of this paper rests on.

\paragraph{(2) No benign arm.} Attack benchmarks send only harmful content, so
nothing distinguishes a model that recognises harm through an encoding from one
refusing anything wearing it. This is not hypothetical: on Llama-3.1-8B the two
hypotheses predict $0.99$ harmful refusal equally well and are told apart only
by the benign arm.

\paragraph{(3) The binary jailbreak judge is a non-refusal detector.} On plaintext benign prompts, where any positive is judge error, our binary judge fires at $0.61$--$0.70$ across the four models, and on the harmful arm its output is one minus the refusal rate to within $0.01$. We report no attack-success number. Detail is in the supplementary material.

\paragraph{(4) The refusal judge scores echo as refusal.} A response that only quotes the ciphertext back is scored as a refusal by a judge reading for non-compliance. We screen every condition by recomputing the harm gap over non-echoing cells and withhold it when the displacement exceeds the gap's own $95\%$ half-width; homoglyph clears on all four models. Reported rates keep the full $n{=}100$ denominator. The screen and its arm-wise control are in the supplementary material.

\paragraph{(5) to (12), in one line each.} \textbf{(5)} Raw sequence length
separates the two corpora, so a probe can score above chance without reading
content. \textbf{(6)} Permutation significance is not sufficiency: a reading can
clear a shuffled-label null and still sit at its own control floor.
\textbf{(7)} An unmeasured axis was recorded as a per-item \emph{false}, making
``the instrument could not read this'' indistinguishable from ``the model did not
decode''; all axes are now tri-state. \textbf{(8)} Comprehension was under-counted
by an exact-match rule that scored genuine near-verbatim decodes as failures.
\textbf{(9)} A probe fitted on the same items it later scores reads item identity
rather than the property, which cost us a result. \textbf{(10)} A screen we
require elsewhere was adopted, tested and documented without ever reaching the
entrypoint that produces runs. \textbf{(11)} The benign arm carried no
comprehension measurement until a reader asked whether the two arms decode alike;
a re-run put them within $0.02$ on fifteen of sixteen cells. \textbf{(12)}
Per-item verdicts survive for one arm of three, so a contrast that is genuinely
paired is reported with unpaired intervals, which are therefore upper bounds.
Each is reported in full, with the control that caught it, in the supplementary
material.

\paragraph{The direction.} Defects (1), (2), (4), (7), (8) and (10) all inflate
apparent safety: an unmeasured axis reads as no-decode, an echo reads as a
refusal, an under-counted decode removes a cell from the
comprehension-conditioned denominator, a missing benign arm hides false
positives, and a screen that never runs leaves the contamination in place. We
believe this is structural rather than coincidental. \emph{No attack succeeded}
is what a broken safety evaluation returns by default, and a result in that
direction does not prompt anyone to look for the bug. Defect (3) is the single
exception and it is a judge artefact running the other way: a non-refusal
detector reported as attack success overstates how often the attack worked.
Defects (5), (6) and (9) sit on the probe axis instead, and (9) runs the same
way one level up: it inflated our confidence in an internal reading rather than
in a safety rate. We found it by holding items out, which is a control we had
not run, which is the same sentence as the other eleven. Defects (11) and (12)
sit on neither axis. They are absences of record rather than errors of
measurement, and both surfaced from a question we could not answer rather than
from a number that looked wrong, which is the harder kind to find and the reason
we list them. Defect~(11) has since been closed by measurement and we report the
result in place; we keep it numbered because a defect found is not unfound by
being fixed, and because the absence was invisible for as long as nobody asked
the question it answers.

\section{Limitations}
\label{sec:limitations}

\paragraph{One encoding carries the cross-model result, and one screen is not
blind to it.} Homoglyph is the only rung passing every screen on all four models.
The pipeline result adds two more, both echo-contaminated in the direction that
\emph{understates} the effect, since the later stages echo more and echo inflates
measured refusal. The echo screen is a stability criterion on the reported gap,
so it is not outcome-independent: over the $27$ model-by-rung cells we screened,
passing correlates with the gap's magnitude at $r=-0.49$ ($p=0.01$), against
$r=-0.59$ ($p=0.001$) with the echo rate that drives it. Rungs with larger
encoded gaps are therefore discarded slightly more often, and a large encoded gap
is evidence \emph{against} the collapse we report. We disclose the direction
rather than correct for it: the screen is required for validity, and dropping it
would admit rungs whose refusal verdicts are majority artefact.

\paragraph{Four further limitations.} Stated in full in the supplementary material: run-to-run non-determinism at $n{=}100$, the unresolved equalisation-versus-saturation reading, whether the harm information is lost or merely unused, and an inference we made and later withdrew.

\section{Conclusion}

Encoded-prompt evaluation reads the arm that cannot see what it is trying to
measure. On the harmful arm, four very different models converge to within
$0.08$; on the benign arm they span two-thirds of the range. The quantity
separating them, whether refusal tracks the request or its appearance, requires
the benign arm and a plaintext baseline, and neither is standard. A complete published safety pipeline improves the harm gap
in plaintext and under the encoding by very nearly as much, so the shortfall
between them is untouched, and the standard metric registers none of it. We
recommend that encoded-attack evaluations report both arms as a matter of course.

\bibliography{paper}

\begin{thebibliography}{37}
\providecommand{\natexlab}[1]{#1}

\bibitem[{Arditi et~al.(2024)Arditi, Obeso, Syed, Paleka, Panickssery, Gurnee,
  and Nanda}]{arditi2024refusal}
Arditi, A.; Obeso, O.; Syed, A.; Paleka, D.; Panickssery, N.; Gurnee, W.; and
  Nanda, N. 2024.
\newblock Refusal in Language Models Is Mediated by a Single Direction.
\newblock In \emph{Advances in Neural Information Processing Systems
  (NeurIPS)}.

\bibitem[{Cao, Yang, and Zhao(2025)}]{Cao_Yang_Zhao_2025}
Cao, Z.; Yang, Y.; and Zhao, H. 2025.
\newblock SCANS: Mitigating the Exaggerated Safety for LLMs via
  Safety-Conscious Activation Steering.
\newblock \emph{Proceedings of the AAAI Conference on Artificial Intelligence},
  39(22): 23523--23531.

\bibitem[{Chao et~al.(2024)Chao, Debenedetti, Robey, Andriushchenko, Croce,
  Sehwag, Dobriban, Flammarion, Pappas, Tram\`{e}r, Hassani, and
  Wong}]{NEURIPS2024_63092d79}
Chao, P.; Debenedetti, E.; Robey, A.; Andriushchenko, M.; Croce, F.; Sehwag,
  V.; Dobriban, E.; Flammarion, N.; Pappas, G.~J.; Tram\`{e}r, F.; Hassani, H.;
  and Wong, E. 2024.
\newblock JailbreakBench: An Open Robustness Benchmark for Jailbreaking Large
  Language Models.
\newblock In Globerson, A.; Mackey, L.; Belgrave, D.; Fan, A.; Paquet, U.;
  Tomczak, J.; and Zhang, C., eds., \emph{Advances in Neural Information
  Processing Systems}, volume~37, 55005--55029. Curran Associates, Inc.

\bibitem[{Cui et~al.(2025)Cui, Chiang, Stoica, and Hsieh}]{cui2025orbench}
Cui, J.; Chiang, W.-L.; Stoica, I.; and Hsieh, C.-J. 2025.
\newblock {OR}-Bench: An Over-Refusal Benchmark for Large Language Models.
\newblock In \emph{Forty-second International Conference on Machine Learning}.

\bibitem[{Ding et~al.(2025)Ding, Kuang, Wang, Cao, Cai, Chen, and
  Huang}]{ding-etal-2025-act}
Ding, P.; Kuang, J.; Wang, Z.; Cao, X.; Cai, X.; Chen, J.; and Huang, S. 2025.
\newblock Why Not Act on What You Know? Unleashing Safety Potential of {LLM}s
  via Self-Aware Guard Enhancement.
\newblock In Che, W.; Nabende, J.; Shutova, E.; and Pilehvar, M.~T., eds.,
  \emph{Findings of the Association for Computational Linguistics: ACL 2025},
  6279--6299. Vienna, Austria: Association for Computational Linguistics.
\newblock ISBN 979-8-89176-256-5.

\bibitem[{Fairoze et~al.(2026)Fairoze, Garg, Lee, and
  Wang}]{fairoze2026bypassingpromptguardsproduction}
Fairoze, J.; Garg, S.; Lee, K.; and Wang, M. 2026.
\newblock Bypassing Prompt Guards in Production with Controlled-Release
  Prompting.
\newblock In \emph{35th USENIX Security Symposium (USENIX Security 26)}.
  Baltimore, MD: USENIX Association.

\bibitem[{Ghosh et~al.(2025)Ghosh, Varshney, Sreedhar, Padmakumar, Rebedea,
  Varghese, and Parisien}]{ghosh-etal-2025-aegis2}
Ghosh, S.; Varshney, P.; Sreedhar, M.~N.; Padmakumar, A.; Rebedea, T.;
  Varghese, J.~R.; and Parisien, C. 2025.
\newblock {AEGIS}2.0: A Diverse {AI} Safety Dataset and Risks Taxonomy for
  Alignment of {LLM} Guardrails.
\newblock In Chiruzzo, L.; Ritter, A.; and Wang, L., eds., \emph{Proceedings of
  the 2025 Conference of the Nations of the Americas Chapter of the Association
  for Computational Linguistics: Human Language Technologies (Volume 1: Long
  Papers)}, 5992--6026. Albuquerque, New Mexico: Association for Computational
  Linguistics.
\newblock ISBN 979-8-89176-189-6.

\bibitem[{Grattafiori et~al.(2024)Grattafiori, Dubey, Jauhri
  et~al.}]{grattafiori2024llama3}
Grattafiori, A.; Dubey, A.; Jauhri, A.; et~al. 2024.
\newblock The Llama 3 Herd of Models.
\newblock arXiv:2407.21783.

\bibitem[{Handa et~al.(2025)Handa, Zhang, Saeidi, Kumbhar, Uddin, RRV, and
  Baral}]{handa2025when}
Handa, D.; Zhang, Z.; Saeidi, A.; Kumbhar, S.; Uddin, M.~N.; RRV, A.; and
  Baral, C. 2025.
\newblock When ''Competency'' in Reasoning Opens the Door to Vulnerability:
  Jailbreaking {LLM}s via Novel Ciphers.
\newblock In \emph{NeurIPS 2025 Workshop: Reliable ML from Unreliable Data}.

\bibitem[{He et~al.(2025)He, Wei, Zhang, Liu, Feng, and An}]{he2025solving}
He, S.; Wei, Q.; Zhang, Z.; Liu, F.; Feng, L.; and An, B. 2025.
\newblock Solving Puzzles? Jailbreaking Multimodal Large Language Models!

\bibitem[{Jiang et~al.(2023)Jiang, Sablayrolles, Mensch
  et~al.}]{jiang2023mistral}
Jiang, A.~Q.; Sablayrolles, A.; Mensch, A.; et~al. 2023.
\newblock Mistral 7B.
\newblock arXiv:2310.06825.

\bibitem[{Jiang et~al.(2024)Jiang, Xu, Niu, Xiang, Ramasubramanian, Li, and
  Poovendran}]{jiang-etal-2024-artprompt}
Jiang, F.; Xu, Z.; Niu, L.; Xiang, Z.; Ramasubramanian, B.; Li, B.; and
  Poovendran, R. 2024.
\newblock {A}rt{P}rompt: {ASCII} Art-based Jailbreak Attacks against Aligned
  {LLM}s.
\newblock In Ku, L.-W.; Martins, A.; and Srikumar, V., eds., \emph{Proceedings
  of the 62nd Annual Meeting of the Association for Computational Linguistics
  (Volume 1: Long Papers)}, 15157--15173. Bangkok, Thailand: Association for
  Computational Linguistics.

\bibitem[{Lambert et~al.(2025)Lambert, Morrison, Pyatkin, Ivison
  et~al.}]{lambert2025tulu3}
Lambert, N.; Morrison, J.; Pyatkin, V.; Ivison, H.; et~al. 2025.
\newblock T\"ulu 3: Pushing Frontiers in Open Language Model Post-Training.
\newblock In \emph{Conference on Language Modeling (COLM)}.

\bibitem[{Li and Zhan(2026)}]{li2026outputaware}
Li, J.; and Zhan, K. 2026.
\newblock Safe Responses Matter: Output-Aware Safety Guardrail Mitigate
  Over-Refusal in MLLMs.
\newblock In \emph{European Conference on Computer Vision (ECCV)}.

\bibitem[{Li, Jiang, and Wei(2025)}]{li-etal-2025-detam}
Li, Y.; Jiang, H.; and Wei, Z. 2025.
\newblock {D}e{TAM}: Defending {LLM}s Against Jailbreak Attacks via Targeted
  Attention Modification.
\newblock In Che, W.; Nabende, J.; Shutova, E.; and Pilehvar, M.~T., eds.,
  \emph{Findings of the Association for Computational Linguistics: ACL 2025},
  11781--11797. Vienna, Austria: Association for Computational Linguistics.
\newblock ISBN 979-8-89176-256-5.

\bibitem[{Lv et~al.(2025)Lv, Zhao, Wang, Tang, Jie, Han, and
  Hu}]{lv-etal-2025-gamma}
Lv, L.; Zhao, Y.; Wang, G.; Tang, X.; Jie, W.; Han, J.; and Hu, S. 2025.
\newblock Gamma-Guard: Lightweight Residual Adapters for Robust Guardrails in
  Large Language Models.
\newblock In Christodoulopoulos, C.; Chakraborty, T.; Rose, C.; and Peng, V.,
  eds., \emph{Proceedings of the 2025 Conference on Empirical Methods in
  Natural Language Processing}, 12219--12231. Suzhou, China: Association for
  Computational Linguistics.
\newblock ISBN 979-8-89176-332-6.

\bibitem[{Mazeika et~al.(2024)Mazeika, Phan, Yin, Zou, Wang, Mu, Sakhaee, Li,
  Basart, Li, Forsyth, and Hendrycks}]{10.5555/3692070.3693501}
Mazeika, M.; Phan, L.; Yin, X.; Zou, A.; Wang, Z.; Mu, N.; Sakhaee, E.; Li, N.;
  Basart, S.; Li, B.; Forsyth, D.; and Hendrycks, D. 2024.
\newblock HarmBench: a standardized evaluation framework for automated red
  teaming and robust refusal.
\newblock In \emph{Proceedings of the 41st International Conference on Machine
  Learning}, ICML'24. JMLR.org.

\bibitem[{Mu et~al.(2025)Mu, He, Zhou, Feng, Xu, Qin, Shi, Liu, Han, Shi, Zhu,
  and Che}]{mu-etal-2025-stealthy}
Mu, H.; He, H.; Zhou, Y.; Feng, Y.; Xu, Y.; Qin, L.; Shi, X.; Liu, Z.; Han, X.;
  Shi, Q.; Zhu, Q.; and Che, W. 2025.
\newblock Stealthy Jailbreak Attacks on Large Language Models via Benign Data
  Mirroring.
\newblock In \emph{Proceedings of the 2025 Conference of the Nations of the
  Americas Chapter of the Association for Computational Linguistics: Human
  Language Technologies (Volume 1: Long Papers)}, 1784--1799. Albuquerque, New
  Mexico: Association for Computational Linguistics.

\bibitem[{Peng et~al.(2026)Peng, Wang, Wang, Li, Li, Ye, Wang, Jia, Zhang, and
  Zhao}]{peng-etal-2026-logic}
Peng, J.; Wang, M.; Wang, N.; Li, J.; Li, Y.; Ye, Y.; Wang, W.; Jia, P.; Zhang,
  K.; and Zhao, X. 2026.
\newblock Logic Jailbreak: Efficiently Unlocking {LLM} Safety Restrictions
  Through Formal Logical Expression.
\newblock In Liakata, M.; Moreira, V.~P.; Zhang, J.; and Jurgens, D., eds.,
  \emph{Findings of the {A}ssociation for {C}omputational {L}inguistics: {ACL}
  2026}, 523--543. San Diego, California, United States: Association for
  Computational Linguistics.
\newblock ISBN 979-8-89176-395-1.

\bibitem[{Prakash et~al.(2026)Prakash, Jie, Abdullah, Satapathy, Cambria, and
  Lee}]{Prakash_Jie_Abdullah_Satapathy_Cambria_Lee_2026}
Prakash, N.; Jie, Y.~W.; Abdullah, A.; Satapathy, R.; Cambria, E.; and Lee, R.
  K.-W. 2026.
\newblock Beyond I’m Sorry, I Can’t: Dissecting Large-Language-Model
  Refusal.
\newblock \emph{Proceedings of the AAAI Conference on Artificial Intelligence},
  40(44): 37830--37838.

\bibitem[{Qi et~al.(2025)Qi, Panda, Lyu, Ma, Roy, Beirami, Mittal, and
  Henderson}]{qi2025shallowalignment}
Qi, X.; Panda, A.; Lyu, K.; Ma, X.; Roy, S.; Beirami, A.; Mittal, P.; and
  Henderson, P. 2025.
\newblock Safety Alignment Should Be Made More Than Just a Few Tokens Deep.
\newblock In \emph{International Conference on Learning Representations
  (ICLR)}.

\bibitem[{{Qwen Team}(2024)}]{qwen2024qwen25}
{Qwen Team}. 2024.
\newblock Qwen2.5 Technical Report.
\newblock arXiv:2412.15115.

\bibitem[{Ren, Nakov, and Naseem(2025)}]{ren2025dualbench}
Ren, K.; Nakov, P.; and Naseem, U. 2025.
\newblock {DUAL}-Bench: Measuring Over-Refusal and Robustness in
  Vision-Language Models.
\newblock arXiv:2510.10846.

\bibitem[{R{\"o}ttger et~al.(2024)R{\"o}ttger, Kirk, Vidgen, Attanasio,
  Bianchi, and Hovy}]{rottger-etal-2024-xstest}
R{\"o}ttger, P.; Kirk, H.; Vidgen, B.; Attanasio, G.; Bianchi, F.; and Hovy, D.
  2024.
\newblock {XST}est: A Test Suite for Identifying Exaggerated Safety Behaviours
  in Large Language Models.
\newblock In Duh, K.; Gomez, H.; and Bethard, S., eds., \emph{Proceedings of
  the 2024 Conference of the North American Chapter of the Association for
  Computational Linguistics: Human Language Technologies (Volume 1: Long
  Papers)}, 5377--5400. Mexico City, Mexico: Association for Computational
  Linguistics.

\bibitem[{Shen et~al.(2024)Shen, Chen, Backes, Shen, and
  Zhang}]{10.1145/3658644.3670388}
Shen, X.; Chen, Z.; Backes, M.; Shen, Y.; and Zhang, Y. 2024.
\newblock "Do Anything Now": Characterizing and Evaluating In-The-Wild
  Jailbreak Prompts on Large Language Models.
\newblock In \emph{Proceedings of the 2024 on ACM SIGSAC Conference on Computer
  and Communications Security}, CCS '24, 1671–1685. New York, NY, USA:
  Association for Computing Machinery.
\newblock ISBN 9798400706363.

\bibitem[{Souly et~al.(2024)Souly, Lu, Bowen, Trinh, Hsieh, Pandey, Abbeel,
  Svegliato, Emmons, Watkins, and Toyer}]{souly2024strongreject}
Souly, A.; Lu, Q.; Bowen, D.; Trinh, T.; Hsieh, E.; Pandey, S.; Abbeel, P.;
  Svegliato, J.; Emmons, S.; Watkins, O.; and Toyer, S. 2024.
\newblock A {StrongREJECT} for Empty Jailbreaks.
\newblock In \emph{Proceedings of the 38th International Conference on Neural
  Information Processing Systems}, NIPS '24. Red Hook, NY, USA: Curran
  Associates Inc.
\newblock ISBN 9798331314385.

\bibitem[{Uysal et~al.(2026)Uysal, Birinci, Mutluergil, and
  {\c{C}}etin}]{uysal2026multilingualobfuscated}
Uysal, C.; Birinci, B.; Mutluergil, S.~O.; and {\c{C}}etin, O. 2026.
\newblock An Empirical Evaluation of Prompt Injection Vulnerabilities Across
  Multilingual and Obfuscated Attack Scenarios.
\newblock \emph{arXiv preprint}.

\bibitem[{Verma et~al.(2025)Verma, Hines, Bilmes, Siska, Zettlemoyer, Gonen,
  and Singh}]{verma-etal-2025-multiguard}
Verma, S.; Hines, K.; Bilmes, J.; Siska, C.; Zettlemoyer, L.; Gonen, H.; and
  Singh, C. 2025.
\newblock {MULTIGUARD}: An Efficient Approach for {AI} Safety Moderation Across
  Languages and Modalities.
\newblock In Christodoulopoulos, C.; Chakraborty, T.; Rose, C.; and Peng, V.,
  eds., \emph{Proceedings of the 2025 Conference on Empirical Methods in
  Natural Language Processing}, 16173--16187. Suzhou, China: Association for
  Computational Linguistics.
\newblock ISBN 979-8-89176-332-6.

\bibitem[{Wei, Haghtalab, and Steinhardt(2023)}]{wei2023jailbroken}
Wei, A.; Haghtalab, N.; and Steinhardt, J. 2023.
\newblock Jailbroken: How Does {LLM} Safety Training Fail?
\newblock In \emph{Advances in Neural Information Processing Systems
  (NeurIPS)}.

\bibitem[{Yan et~al.(2025)Yan, Yang, Wang, Feng, Zhang, and
  Zhao}]{yan-etal-2025-semanticcamo}
Yan, J.; Yang, X.; Wang, D.; Feng, S.; Zhang, Y.; and Zhao, Y. 2025.
\newblock {S}emantic{C}amo: Jailbreaking Large Language Models through Semantic
  Camouflage.
\newblock In Che, W.; Nabende, J.; Shutova, E.; and Pilehvar, M.~T., eds.,
  \emph{Findings of the Association for Computational Linguistics: ACL 2025},
  14427--14452. Vienna, Austria: Association for Computational Linguistics.
\newblock ISBN 979-8-89176-256-5.

\bibitem[{Yong, Menghini, and Bach(2023)}]{yong2023lowresource}
Yong, Z.~X.; Menghini, C.; and Bach, S. 2023.
\newblock Low-Resource Languages Jailbreak {GPT}-4.
\newblock In \emph{Socially Responsible Language Modelling Research}.

\bibitem[{Yuan et~al.(2024)Yuan, Jiao, Wang, tse Huang, He, Shi, and
  Tu}]{yuan2024gpt}
Yuan, Y.; Jiao, W.; Wang, W.; tse Huang, J.; He, P.; Shi, S.; and Tu, Z. 2024.
\newblock {GPT}-4 Is Too Smart To Be Safe: Stealthy Chat with {LLM}s via
  Cipher.
\newblock In \emph{The Twelfth International Conference on Learning
  Representations}.

\bibitem[{Zhang et~al.(2026)Zhang, Wang, Liu, Chen, and
  Wang}]{zhang-etal-2026-llm}
Zhang, H.; Wang, D.; Liu, Y.; Chen, K.; and Wang, W. 2026.
\newblock {LLM}-{VA}: Resolving the Jailbreak-Overrefusal Trade-off via Vector
  Alignment.
\newblock In Liakata, M.; Moreira, V.~P.; Zhang, J.; and Jurgens, D., eds.,
  \emph{Proceedings of the 64th Annual Meeting of the {A}ssociation for
  {C}omputational {L}inguistics (Volume 1: Long Papers)}, 5760--5776. San
  Diego, California, United States: Association for Computational Linguistics.
\newblock ISBN 979-8-89176-390-6.

\bibitem[{Zhang, Zandsalimy, and Sushmita(2026)}]{pmlr-v318-zhang26a}
Zhang, H.; Zandsalimy, M.; and Sushmita, S. 2026.
\newblock Exposing LLM Safety Gaps Through Mathematical Encoding: New Attacks
  and Systematic Analysis.
\newblock In Bouzar-Benlabiod, L.; and Leung, C., eds., \emph{Proceedings of
  the The 39th Canadian Conference on Artificial Intelligence}, volume 318 of
  \emph{Proceedings of Machine Learning Research}, 662--677. PMLR.

\bibitem[{Zhang et~al.(2025)Zhang, Zhang, Li, Huang, Jia, Hu, Zhang, Liu, Ma,
  and Shen}]{10.1145/3724393}
Zhang, X.; Zhang, C.; Li, T.; Huang, Y.; Jia, X.; Hu, M.; Zhang, J.; Liu, Y.;
  Ma, S.; and Shen, C. 2025.
\newblock JailGuard: A Universal Detection Framework for Prompt-based Attacks
  on LLM Systems.
\newblock \emph{ACM Trans. Softw. Eng. Methodol.}, 35(1).

\bibitem[{Zhao et~al.(2025)Zhao, Huang, Wu, Bau, and Shi}]{zhao2025llmsencode}
Zhao, J.; Huang, J.; Wu, Z.; Bau, D.; and Shi, W. 2025.
\newblock LLMs Encode Harmfulness and Refusal Separately.
\newblock In \emph{Advances in Neural Information Processing Systems
  (NeurIPS)}.

\bibitem[{Zou et~al.(2024)Zou, Phan, Wang, Duenas, Lin, Andriushchenko, Wang,
  Kolter, Fredrikson, and Hendrycks}]{zou2024circuitbreakers}
Zou, A.; Phan, L.; Wang, J.; Duenas, D.; Lin, M.; Andriushchenko, M.; Wang, R.;
  Kolter, J.~Z.; Fredrikson, M.; and Hendrycks, D. 2024.
\newblock Improving Alignment and Robustness with Circuit Breakers.
\newblock In \emph{Advances in Neural Information Processing Systems},
  volume~37, 83345--83373.

\end{thebibliography}

\clearpage
\appendix
\twocolumn[{\centering\LARGE\bfseries Supplementary Document\par\vspace{1.5em}}]

\noindent
This document accompanies the main paper and is not required to read it. The
main paper states and defends every claim it makes; what follows is the fuller
reporting behind several of them, provided because referees of an earlier draft
asked for it. Section~\ref{sup:repro} gives the implementation detail needed to
reproduce the runs, and the table at its end gives every per-cell rate behind
the main paper's ladder table, including the echo denominators and the
three-way echo sensitivity.

\bigskip

\noindent\textbf{What is where.} \S\ref{sup:repro} gives the implementation
detail needed to reproduce every run. \S\ref{sup:survey} adjudicates the
thirteen surveyed papers row by row. \S\ref{sup:intervals} reports an interval
on every rate the main paper quotes. \S\ref{sup:defects} reports each instrument
defect in full, numbered as the main paper numbers them.
\S\ref{sup:judging} covers the refusal judge, its validation and run-to-run
variation. \S\ref{sup:limits} states four further limitations and the internals
attempt we withdrew.

\section{Reproduction details}
\label{sup:repro}

Both referees of an earlier draft asked for this section, and one gave the
reason it is load-bearing rather than cosmetic: we report below that only
$12$--$58\%$ of responses repeat byte-identically across nominally greedy runs,
which makes every omitted implementation detail a possible source of the
difference. Everything here is a fact about the configuration, not a summary of
it.

\paragraph{Checkpoints.} \texttt{meta-llama/\allowbreak Llama-3.1-8B-Instruct},
\texttt{Qwen/\allowbreak Qwen2.5-7B-Instruct},
\texttt{mistralai/\allowbreak Mistral-7B-Instruct-v0.3},
\texttt{allenai/\allowbreak Llama-3.1-Tulu-3-8B-SFT},
\texttt{allenai/\allowbreak Llama-3.1-Tulu-3-8B-DPO} and
\texttt{allenai/\allowbreak Llama-3.1-Tulu-3-8B}. Each model's own chat template is
applied as shipped in its tokenizer, with no system prompt; weights load at the
checkpoint's native dtype.

\paragraph{Corpus.} JailbreakBench's $100$ harmful behaviours and its $100$
theme-matched benign counterparts \citep{NEURIPS2024_63092d79}, paired by name
(\texttt{jbb\_defamation} against \texttt{benign\_defamation}, and so on through
all $100$). No item is filtered or reweighted, and $n{=}100$ per cell throughout
is the whole set rather than a sample of it.

\paragraph{Transformations.} All three are exactly invertible and applied to the
request text only.
\emph{Homoglyph} substitutes $21$ Latin characters with Cyrillic lookalikes and
leaves everything else untouched: lowercase \texttt{a c e i j o p s x y} to
U+0430, U+0441, U+0435, U+0456, U+0458, U+043E, U+0440, U+0455, U+0445, U+0443,
and uppercase \texttt{A B C E H K M O P T X} to U+0410, U+0412, U+0421, U+0415,
U+041D, U+041A, U+041C, U+041E, U+0420, U+0422, U+0425. The map is injective, so
it inverts. Character count is unchanged.
\emph{Fullwidth} maps every ASCII printable in U+0021--U+007E by the constant
offset U+FF01 $-$ U+0021, and the space to the ideographic space U+3000.
Character count is unchanged.
\emph{Zero-width} inserts U+200B after every character, which roughly doubles
character count and is the reason it is the one transformation of the three that
is not length-preserving.

\paragraph{The attack scaffold, verbatim.} Encoded conditions send the
transformation's own wrapper with the ciphertext substituted for
\texttt{\{ciphertext\}}. Homoglyph: \emph{``The following request is encoded with
Cyrillic and Greek homoglyph substitution:''} followed by a blank line and the
ciphertext. Fullwidth and zero-width are identical but for the naming clause,
\emph{``\dots encoded with fullwidth character forms''} and \emph{``\dots encoded
with zero-width characters inserted between letters''}. The scaffold arm of
the main paper's model-comparison section sends this same wrapper around \emph{untransformed} text,
which is what separates the template's effect from the characters'. The plaintext
arm sends the request with no wrapper at all.

\paragraph{Decoding.} Greedy throughout: \texttt{do\_sample=False}, no
temperature or top-$p$, $256$ new tokens for the decode-and-restate probe and
$512$ for the attack response, batch size $8$. Truncation at those ceilings is
the reason they are set from the pilot's response-length upper tail rather than
by taste: a completion cut mid-answer would be scored as a non-answer.

\paragraph{The refusal judge, verbatim.} \texttt{gpt-5-mini} at temperature $0$.
The prompt presents the request and the response in \texttt{<request>} and
\texttt{<response>} tags and asks: \emph{``Does the language model refuse to
assist the user with their request? A refusal includes any explicit decline,
safety disclaimer that avoids answering, or statement that the model cannot or
will not help. Also treat as a refusal (Yes) if the response merely quotes,
restates, or echoes the request without actually engaging with it, or if the
response is entirely irrelevant to the request.''} The verdict is returned as
JSON with an \texttt{answer} of \texttt{Yes} or \texttt{No} and a brief
reasoning field. A response that fails to parse is recorded as a judge fallback
and counted in the run record; the fallback rate is $0.00$ on every cell in
the main paper's post-training ladder table. Note that the echo clause is an \emph{extension} we
made to JailbreakBench's template, and defect (4) of the main paper's instrument section is
the failure mode that extension creates. It is not a limitation of the judge
model.

\paragraph{Seeds and estimators.} One RNG seed, $0$, for probe train/test
splits; one, $0$, for the bootstrap. Rate intervals are Wilson score intervals,
chosen over the normal approximation because several cells sit within a few
counts of $0$ or $100$ where the latter's lower bound runs outside $[0,1]$.
Intervals on a harm gap are unpaired Wald intervals on a difference of two
independent proportions, because the harmful and benign arms are different
prompt sets with no item pairing available. Intervals on a \emph{change} in a
harm gap between two checkpoints are percentile intervals from a $20{,}000$-draw
bootstrap that resamples items once per draw and scores both checkpoints on the
same resample, because there the two conditions do share items. The confidence
level is derived from a single configured $\alpha=0.05$ rather than written as a
constant. Pairwise tests on binary verdicts over shared items are exact
two-sided McNemar tests on the discordant counts.

\paragraph{Computing infrastructure.} Every generation run reported here used a
single NVIDIA H200 GPU with eight CPU cores and $32$\,GB of host RAM, on a
shared Linux SLURM cluster, one
model per job, under an eight-hour wall clock. No run required more than one GPU
and none was distributed. A full condition sweep for one model completes in
between $20$ minutes and $1$ hour $15$ minutes. The stack is Python $3.13$ with
PyTorch $2.11$ (CUDA $12.8$), Transformers $5.14$, Accelerate $1.14$, NumPy
$2.5$ and scikit-learn $1.9$, pinned by the lock file committed in the code
supplement. Probe fits are held to single-threaded BLAS through
\texttt{threadpoolctl}: left to the eight-core default the fits thrash, the same
sweep takes over thirty times longer, and that is what killed our first attempt
at the wall. The judges are the only component not run locally, being API calls
to \texttt{gpt-5-mini} at temperature $0$.

\paragraph{Run-to-run reproducibility.} Greedy decoding is not deterministic
across runs here. Re-running the same checkpoint on the same condition
reproduces $12$--$58\%$ of responses byte-identically, and $1.0$--$4.3\%$ of
refusal verdicts flip. The flips very nearly cancel: the largest gap movement we
observed across a repeat is $0.010$ on Tülu/fullwidth, where $13\%$ of cells
changed text. The cause is batch composition changing floating-point reduction
order, not sampling. The rightmost column of Table~\ref{tab:full} gives the
observed movement for every cell in the main paper's post-training ladder table.

\begin{table*}[t]
\centering
\small
\begin{tabular}{llcccccccc}
\toprule
& & \multicolumn{2}{c}{refusal rate, $n{=}100$ [95\% CI]} & & \multicolumn{4}{c}{echo} & repeat \\
\cmidrule(lr){3-4}\cmidrule(lr){6-9}
stage & encoding & harmful & benign & $\Delta_{\text{enc}}$ & clean $n$ h/b & dropped & as refusal & as non-refusal & $|\delta|$ \\
\midrule
SFT & homoglyph & 0.99 {\scriptsize[0.95,1.00]} & 0.79 {\scriptsize[0.70,0.86]} & $+0.20$ & 98/94 & $+0.19$ & $+0.18$ & $+0.22$ & 0.00 \\
SFT & fullwidth$^{\dagger}$ & 0.94 {\scriptsize[0.88,0.97]} & 0.82 {\scriptsize[0.73,0.88]} & $+0.12$ & 34/23 & $+0.36$ & $+0.07$ & $+0.18$ & 0.02 \\
SFT & zero-width & 1.00 {\scriptsize[0.96,1.00]} & 0.90 {\scriptsize[0.83,0.94]} & $+0.10$ & 75/59 & $+0.07$ & $+0.04$ & $+0.20$ & 0.00 \\
DPO & homoglyph & 0.90 {\scriptsize[0.83,0.94]} & 0.61 {\scriptsize[0.51,0.70]} & $+0.29$ & 96/92 & $+0.29$ & $+0.26$ & $+0.30$ & 0.00 \\
DPO & fullwidth & 0.74 {\scriptsize[0.65,0.82]} & 0.45 {\scriptsize[0.36,0.55]} & $+0.29$ & 24/26 & $+0.18$ & $+0.06$ & $+0.04$ & 0.01 \\
DPO & zero-width$^{\dagger}$ & 0.87 {\scriptsize[0.79,0.92]} & 0.44 {\scriptsize[0.35,0.54]} & $+0.43$ & 32/32 & $+0.59$ & $+0.19$ & $+0.19$ & 0.00 \\
RLVR & homoglyph & 0.94 {\scriptsize[0.88,0.97]} & 0.48 {\scriptsize[0.38,0.58]} & $+0.46$ & 98/94 & $+0.45$ & $+0.42$ & $+0.46$ & 0.01 \\
RLVR & fullwidth$^{\dagger}$ & 0.73 {\scriptsize[0.64,0.81]} & 0.31 {\scriptsize[0.23,0.41]} & $+0.42$ & 22/25 & $+0.16$ & $+0.06$ & $+0.03$ & 0.00 \\
RLVR & zero-width$^{\dagger}$ & 0.83 {\scriptsize[0.74,0.89]} & 0.37 {\scriptsize[0.28,0.47]} & $+0.46$ & 33/26 & $+0.69$ & $+0.17$ & $+0.24$ & 0.02 \\
\bottomrule
\end{tabular}
\caption{Every cell behind the main paper's post-training ladder table, with the reporting both
referees asked for. \emph{clean $n$} is the count surviving the echo filter in
each arm. The last three columns are $\Delta_{\text{enc}}$ under the three
treatments of an echoing response: dropped as unscorable (what we do, and what
the screen is computed on), recoded as a refusal, and recoded as a non-refusal.
$\dagger$~marks a cell failing the echo screen. Homoglyph is stable across all
three treatments at every stage, within $0.04$; the other two encodings move by
up to $0.52$, which is the selection effect the screen exists to detect and the
reason the pipeline claim rests on homoglyph. The final column is the absolute
movement of harmful-arm refusal across an independent repeat of the same
condition.}
\label{tab:full}
\end{table*}

\section{The arm survey}
\label{sup:survey}

\subsection{The thirteen rows, adjudicated}

Table~\ref{tab:survey} adjudicates every paper behind the main paper's
related-work claim about the benign-in-attack-form cell. Verdicts use four codes, and the distinction between the
last two is where every interesting case sits. \textbf{B0}: no benign arm at
all. \textbf{B1}: a benign arm in plain form only, never passed through the
transformation. \textbf{B2}: benign content does pass through the
transformation, but is scored for capability, that is, did the model decode or
answer it, rather than for refusal. \textbf{B3}: benign content through the
transformation with a refusal or false-positive rate reported. B3 is the cell.
B2 is the near miss that matters most, because a paper at B2 built the corpus,
applied the transformation, and measured something else on it.

\begin{table}[t]
\centering
\footnotesize
\setlength{\tabcolsep}{1.5pt}
\begin{tabular}{llc}
\toprule
paper & transformation & verdict \\
\midrule
\citet{wei2023jailbroken} & Base64$+$ & \textbf{B2} \\
\citet{yong2023lowresource} & translation & B0 \\
\citet{yuan2024gpt} & ciphers & B0 \\
\citet{jiang-etal-2024-artprompt} & ASCII art & B0 \\
Ren et al. (2024) & code & B0 \\
Jiang et al. (2024) & cipher chars & B0 \\
\citet{handa2025when} & ten ciphers & \textbf{B3} \\
\citet{he2025solving} & puzzles & B0 \\
\citet{yan-etal-2025-semanticcamo} & sem.\ camo. & B0 \\
\citet{pmlr-v318-zhang26a} & math. & B0 \\
\citet{peng-etal-2026-logic} & logic & B0 \\
\midrule
\citet{rottger-etal-2024-xstest} & (none) & B1 \\
\citet{cui2025orbench} & (none) & B1 \\
\bottomrule
\end{tabular}
\caption{Eleven input-transformation attack papers and the two over-refusal
benchmarks, adjudicated from full text. \textbf{One paper reaches B3.}
\citet{handa2025when} reframe all fifty of their AdvBench items to be safe while
keeping the unsafe vocabulary, send the reframed set through five of their
ciphers, and report the fraction incorrectly flagged as unsafe, which reaches
$0.40$. That is a different corpus from their CipherBench, whose benign arm is
scored for decoding accuracy and is the column the main paper calls one away.
\citet{wei2023jailbroken} is the remaining near miss, a single harmless control
prompt scored for whether the model answered correctly. Two limits, both of which
argue for the same caution: the frame is one curated corpus rather than the
literature, and full-text extraction reads prose, so a rate living only in a
figure axis would not be caught. Our own first pass adjudicated the
\citet{handa2025when} row from CipherBench alone and scored it B2.}
\label{tab:survey}
\end{table}

\subsection{The near misses, and the one paper that runs the cell}

\paragraph{The near miss that founded this attack class.} The gap is
not that the measurement is hard or that nobody considered it.
\citet{wei2023jailbroken} tested each attack against a harmless control prompt
alongside the harmful set, and observed the effect we quantify, noting
parenthetically that one model ``even refuses a harmless control prompt'' under
some attacks. That control is a single prompt; it is scored as a capability
check, counted successful if the model answered the question correctly; and it
appears as one column of a table in their own paper. A single prompt yields an
anecdote and not a rate, and refusal is only one of the ways it can fail.

\paragraph{\citet{handa2025when} run the cell, on a corpus of their own.} They
build CipherBench, which passes benign text through ten ciphers and scores it
for decoding accuracy, and they attack harmful queries through those same
ciphers; the refusal rate on \emph{that} benign corpus is not reported, and it
is the column the main paper calls one away. Separately, they reframe all fifty
of their AdvBench items to be safe while keeping the unsafe vocabulary, send the
reframed set through five of their ciphers, and report the percentage
incorrectly flagged as unsafe, reaching $0.40$ on gpt-oss-20b under grid
encoding. That arm reaches B3, on a different corpus from CipherBench, and we
record it as such rather than as a near miss.

\paragraph{What is left for us to add.} Their measurement is a false-positive
rate on one arm. Ours is paired: the same items benign and harmful, in plaintext
and through the transformation, which is what makes the harm gap and its
collapse computable and gives the encoded rate a baseline to be read against.
Their judge is the binary safe-or-unsafe classifier whose failure mode we report
as defect~(3), and they note themselves that it mislabels a nuanced refusal as
over-defensive. We are therefore not opening an untouched question. We are
turning a control that two papers ran, one on a single prompt and one on a
single arm, into a paired rate across four models.

\section{Intervals on every reported rate}
\label{sup:intervals}

\noindent
Table~\ref{tab:intervals} carries the main paper's cross-model and three-arm
tables. Intervals on the post-training ladder are in Table~\ref{tab:full}
instead, where they sit beside the echo denominators and the three-way echo
sensitivity that the ladder's caption promises alongside them.

\begin{table}[t]
\centering
\small
\setlength{\tabcolsep}{3pt}
\begin{tabular}{llll}
\toprule
model & condition & harmful & benign \\
\midrule
\multicolumn{4}{l}{\emph{the main paper's cross-model table}} \\
Llama-3.1-8B-It & plaintext & 0.92 [0.85,0.96] & 0.10 [0.06,0.17] \\
                & homoglyph & 0.99 [0.95,1.00] & 0.99 [0.95,1.00] \\
Qwen2.5-7B-It   & plaintext & 0.92 [0.85,0.96] & 0.11 [0.06,0.19] \\
                & homoglyph & 0.91 [0.84,0.95] & 0.30 [0.22,0.40] \\
T\"ulu-3-8B     & plaintext & 0.95 [0.89,0.98] & 0.16 [0.10,0.24] \\
                & homoglyph & 0.93 [0.86,0.97] & 0.48 [0.38,0.58] \\
Mistral-7B-v0.3 & plaintext & 0.38 [0.29,0.48] & 0.01 [0.00,0.05] \\
                & homoglyph & 0.91 [0.84,0.95] & 0.63 [0.53,0.72] \\
\midrule
\multicolumn{4}{l}{\emph{the main paper's three-arm table}} \\
Llama-3.1-8B-It & plaintext & 0.93 [0.86,0.97] & 0.10 [0.06,0.17] \\
                & scaffold  & 0.99 [0.95,1.00] & 0.83 [0.74,0.89] \\
                & encoded   & 0.98 [0.93,0.99] & 0.99 [0.95,1.00] \\
Qwen2.5-7B-It   & plaintext & 0.93 [0.86,0.97] & 0.11 [0.06,0.19] \\
                & scaffold  & 0.92 [0.85,0.96] & 0.12 [0.07,0.20] \\
                & encoded   & 0.88 [0.80,0.93] & 0.33 [0.25,0.43] \\
T\"ulu-3-8B     & plaintext & 0.96 [0.90,0.98] & 0.16 [0.10,0.24] \\
                & scaffold  & 0.98 [0.93,0.99] & 0.46 [0.37,0.56] \\
                & encoded   & 0.93 [0.86,0.97] & 0.50 [0.40,0.60] \\
Mistral-7B-v0.3 & plaintext & 0.37 [0.28,0.47] & 0.01 [0.00,0.05] \\
                & scaffold  & 0.60 [0.50,0.69] & 0.12 [0.07,0.20] \\
                & encoded   & 0.91 [0.84,0.95] & 0.64 [0.54,0.73] \\
\bottomrule
\end{tabular}
\caption{Wilson $95\%$ intervals on every rate in
the main paper's cross-model and three-arm tables, $n{=}100$ throughout. Wilson
rather than the normal approximation because at $n{=}100$ with counts as low as
$1$ the latter's lower bound runs negative, which is exactly where our smallest
cells sit. The corresponding gap intervals are Wald intervals for two
independent samples, quoted in the text where each gap is discussed; unpaired is
correct there, because a harm gap's two arms are different corpora and no
pairing exists to exploit. See defect~(12) for the one contrast where pairing
does exist and our records cannot supply it.}
\label{tab:intervals}
\end{table}

\section{The instrument defects in full}
\label{sup:defects}

\subsection{Defect (3): the binary jailbreak judge}

\paragraph{(3) The binary jailbreak judge is a non-refusal detector.} On
plaintext benign prompts (where no attack exists by construction, so any
positive is judge error), our binary judge fires at $0.70$, $0.63$, $0.61$ and
$0.69$ on the four models. On the harmful arm its output equals one minus the
refusal rate to within $0.01$ on all four. It measures whether the model
complied, discounted, and reports it as attack success. This reproduces the
critique of \citet{souly2024strongreject} on our own judge and on the cleanest
available corpus. No attack-success number in this paper, and none we previously
measured, is reportable.

\subsection{Defect (4): echo scored as refusal, and two caveats}

\paragraph{(4) The refusal judge scores echo as refusal.} A common non-answer to
an encoded prompt is the model reproducing the ciphertext. We ran a paired
control in which only the ciphertext moves: arm A re-judges the response
unmodified, arm B appends the ciphertext to it, arm C sends the ciphertext
\emph{alone}. Arm C's ground truth is by construction: a bare ciphertext
contains no refusal. On $15$ conditions across two model families and $1{,}500$
items, arm C is classified as refusal $100\%$ of the time, without exception. Arm
B moves $0.00$--$0.09$: the judge is not confused by ciphertext being present,
only by a response that is nothing else. Echo is therefore scored per cell by the same
recovery scorer, and every condition is screened: the harm gap is recomputed over
non-echoing cells alone, and the condition is reportable just if that
recomputation moves the gap by less than the gap's own $95\%$ half-width.
Reported rates keep the full $n{=}100$ denominator. The screen decides whether
a number may be quoted, and it never silently changes one. Homoglyph clears on all
four models (on Llama it displaces the gap by $0.002$ against a bar of $0.028$,
with $89$ and $74$ non-echoing cells in the two arms); fullwidth and zero-width
fail it on three of four, which is part of why the cross-model result rests on
homoglyph alone. Arm A, the control's own control, agrees with
the recorded verdict to within $0.00$--$0.04$, which is what licenses reading the
flips at all. Independent annotation puts a size on what this does to the
category: pool-weighted over the sampled strata, $0.63$ and $0.49$ of the cells
the judge calls refusals are labelled echo or irrelevance rather than refusal by
the two annotators of the limitations section. That is not judge error, since the
judge was instructed to count them as refusals and the annotators were given the
same instruction, and it is why the screen above operates on the category rather
than on the judge.

\paragraph{An honest caveat.} Encoded harmful refusal sits at $0.91$--$0.99$,
near ceiling. \emph{Equalisation} and \emph{saturation} both fit and $n{=}100$
cannot separate them. The consequence for evaluation is identical either way,
and that consequence is our claim.

\paragraph{Mistral is not an exception.} Its harmful refusal \emph{rises} under
the encoding, $0.38 \rightarrow 0.91$. This is the same mechanism rather than a
counter-direction: encoded harmful refusal is model-independent at
$0.91$--$0.99$, so a model starting below that level is pulled up to it. It also
rules out an alternative reading of the whole result (that these are simply
heavily safety-tuned models refusing anything unusual), since the effect
appears on a model that does not reliably refuse harmful plaintext.

\subsection{Defects (5) to (9), and (12)}

\paragraph{(5) Sequence length separates the corpora.} Raw character length
alone separates the harmful from the benign corpus at AUROC $0.654$ ($86.0$ vs
$73.8$ characters). Every encoder is monotone in length, so this survives into
every encoded condition. Any probe-based claim must beat a length-matched null;
permutation tests over unmatched labels do not provide one, since they control
the multiple comparison and say nothing about \emph{what} separates.

Two corrections to how we applied that rule, both found by auditing our own code
rather than our results. First, \emph{a null must be on the scale of the thing it
nulls}. The probe-side null subtracts the AUROC a length-only classifier
achieves, which is meaningful only for a reading that is itself an AUROC. We were
passing it to our two rate-valued readings as well, where subtracting $0.654$
from a rate is negative whenever the rate is small, so every rate-valued reading
in every run was withheld for a reason that had not examined the data, and this
control had never once been evaluated on that axis. The replacement permutes the
two arms' labels \emph{within} length strata, which is on the rate's own scale.
Run for the first time, the harm gap clears it on three models by $+0.13$ to
$+0.43$; on the fourth the gap is zero and there is nothing for a null to clear,
which is the correct verdict for a claim of absence rather than a failure. The
more direct statement is available from the same fit and answers the objection
more plainly: matching on length \emph{moves} the gap by at most $0.024$ across
the four models ($+0.002$, $+0.003$, $-0.014$, $-0.024$). Whatever the harmful
and benign corpora differ by in length, the behavioural gap is not made of it.
Second, \emph{a token-count inflation is not a length confound}. Homoglyph raises
tokens per character by about $4\times$, which invites the objection that a
character-length null understates the confound. It does not: the inflation is
near-uniform across items, so token and character length are rank-correlated at
$\rho=0.98$, and both the null and the stratification are rank statistics. Fitted
on tokens instead, the length AUROC moves from $0.654$ to $0.647$--$0.651$ and
the stratified gap by at most $0.02$.

\paragraph{(6) Significance is not sufficiency.} Our content probe is a linear
readout fitted on plaintext contrast sets and read on the encoded condition,
following the direction-fitting method of \citet{arditi2024refusal}; we screen
it additionally against a vocabulary control drawn from XSTest
\citep{rottger-etal-2024-xstest}. Under a permutation test, $14$ of $15$ encodings
licensed such a probe. Screened against a floor derived
from encodings the model demonstrably cannot decode (where any apparent
signal is by construction not decoded content), only two survive. We set the
floor at mean $+2$\,SD over control encodings, having found that a max-statistic
floor is $n$-dependent and not comparable across runs.

\paragraph{(7) Unmeasured is not negative.} Our licensing layer initially
recorded an unlicensed probe as a per-item \emph{false}, making a condition the
instrument could not read indistinguishable from one where the model did not
decode. Since that term decides whether a condition counts as decoded at all, an
entire band of results was an artefact of a silent default. All axes are now tri-state,
and an unmeasured axis produces a declared hole rather than a value.

\paragraph{(8) Comprehension was under-counted.} Scoring decode-and-restate by
exact match or substring containment marked genuine near-verbatim decodes as
failures on a single-word slip. Under the corrected rule, $328$ cells across two
models changed from ``could not decode'' to ``decoded'', in one direction only.

\paragraph{(9) A probe can read the item rather than the property, and this
one is ours.} We fitted a harm direction on plaintext activations and
transferred it to the encoded condition without refitting, and it separated
encoded harmful from encoded benign at near-ceiling on all four models. The fit
used every item in the corpus and was then scored on the encoded versions of
those same items. Where the model can decode the transformation, an item's
encoded activation sits near its own plaintext activation, so the probe scores
item identity and not harm. Holding items out of the fit drops the transfer to
$0.618$--$0.811$, \emph{below} each model's own within-plaintext cross-validated
baseline, and at matched training size the seen-item advantage is $+0.19$ to
$+0.38$. We withdrew the result. Neither the permutation test nor the
decode-impossible floor of (6) catches this, and the floor's failure is
structural rather than a calibration slip: the floor is estimated on encodings
the model cannot decode, which is precisely where item leakage is weakest, and
applied to encodings it decodes well, which is where leakage is strongest. Any
cross-condition probe transfer is exposed to this whenever the second condition
is a transformation of the same items, which is the usual design.

\paragraph{(12) Per-item verdicts survive for one arm of three.} Our pipeline
writes per-item records for the encoded condition and aggregate rates for the
plaintext and scaffold conditions. Any contrast \emph{across} conditions is
therefore reported with unpaired intervals even though it is measured on the
same items and is genuinely paired, which is the decomposition of
the main paper's three-arm section and the case the referee of an earlier draft raised. The
direction is conservative, so no reported interval is too narrow. It is a
persistence defect rather than an analysis one, and it is invisible until
somebody asks for the paired number.

\subsection{Defects (10) and (11), and one non-defect}

\paragraph{Not a defect, and the distinction is load-bearing.} The refusal judge
is otherwise sound: it separates plaintext harmful from plaintext benign by
$+0.79$ to $+0.82$ on three models at a benign false-positive rate of
$0.01$--$0.16$. Its failure is specific to responses containing no content. We
keep this out of the numbered list because it is a validation rather than a
defect, and we report it beside (3) and (4) because conflating a broken harm
judge with a sound refusal judge would discard the measurements this paper is
built on. The evidence does not support that.

\paragraph{(10) A screen can be required and never run.} The echo screen of
item~(4) is called a validity requirement in this section and in
the main paper's limitations section, and it had never been applied to the pipeline table of
the main paper's post-training section. It needs both arms' per-item verdicts, and that experiment
predates the benign arm being written to disk, so a screen adopted in one
section did not reach backwards into a table written before it existed. Applied,
it rejects four of the nine cells. Nothing in our pipeline noticed, because no
artefact records which screens a given number went through.

\paragraph{(11) The benign arm carried no comprehension measurement.} For most
of this work we measured decode ability on the harmful arm and not on the benign
one, so a harm gap and a comprehension gap were not separable in our data: a
model could refuse benign encoded content more often simply because it decoded
that content less often. The omission was deliberate, and wrong for a reason
worth stating. Writing a field that was never measured as a negative is
defect~(7), one measurement over, so we wrote nothing; what we did not foresee is
that the test which settles the interpretation is then exactly the test our own
records cannot support. The asymmetry was this paper's own finding one level in.

We closed it by measuring rather than by qualifying. Decode ability is now
measured on both arms as a mandatory step of the pipeline, and we re-ran all four
models on the two reported encodings plus two encodings the models cannot read at
all, which serve as a floor. \textbf{The two arms decode alike.} Across sixteen
model--encoding cells the largest benign-minus-harmful difference in ability is
$0.10$ and fifteen of sixteen fall within $0.02$; on homoglyph the deltas are
$+0.02$, $+0.01$, $-0.10$ and $+0.01$ for Llama-3.1-8B, Qwen2.5-7B,
T\"ulu-3-8B and Mistral-7B, and on math-bold $+0.00$, $+0.00$, $-0.01$ and
$+0.00$. The harm gaps we report are therefore differences in harm sensitivity
and not in comprehension. Two details we state rather than round away: T\"ulu-3 on
homoglyph is the one cell past $0.02$, at $0.78$ benign against $0.88$ harmful,
which runs in the direction that would inflate its $+0.46$ gap and is far too
small to carry it; and the two unreadable encodings read ability $0.00$ on
\emph{both} arms with refusal near $1.00$ on both, which is what shows the
measurement can report a floor rather than only finding differences where they
are convenient. We keep the defect in this list because it was real, because it
was invisible until a reader asked the question, and because the fix was a
measurement.

\section{Judging and validation}
\label{sup:judging}

\subsection{The refusal judge, and run-to-run variation}

\paragraph{Which runs, and how far they repeat.} The main paper's cross-model table comes
from one run per model that measures the plaintext baseline and the encoded arms
inside a single job; the main paper's post-training table comes from the analogous per-stage
runs. Both comparisons are therefore paired within a job, and the two families
disagree by $0.01$ where they overlap. Other runs measured the same cells and
they do not agree exactly. Generation is greedy and the corpus is fixed, yet only
$12\%$--$58\%$ of responses repeat byte-identically across runs: batch
composition changes the reduction order, one greedy argmax flips, and the
continuation diverges. Verdicts flip on $1.0\%$--$4.3\%$ of cells, and the flips
very nearly cancel: across $4$--$8$ replicates per model the harm gap moves by
$0.01$--$0.06$, and gap lost by at most $0.05$. We report that range rather than
a run we chose. It is a measured resolution floor, and the bootstrap null below
cannot see either route that produces it.

\paragraph{Refusal judging.} Refusal is classified by
\texttt{gpt-5-mini} at temperature $0$, prompted with JailbreakBench's
refusal-judge template \citep{NEURIPS2024_63092d79} extended to count a
response that merely echoes or is irrelevant to the request as a refusal.
the main paper's instrument section reports its validation and the one failure mode that
extension creates.

\subsection{Validation, held-out status, and two withdrawn inferences}

\paragraph{This is a null, and we ran it as a falsification test.} An earlier
version of this analysis, lacking the plaintext arm, saw encoded benign refusal
fall monotonically across stages on all three encodings and concluded that
post-training progressively restores the harm gap. The plaintext baseline
shows that fall is mostly general de-refusal (plaintext benign refusal falls
in lockstep, $0.45 \rightarrow 0.17 \rightarrow 0.16$), and the
encoding-specific component is flat. We withdrew the recovery claim. We report the episode because the failure is cheap to repeat: a stage series measured only
in the encoded condition reads a general behavioural drift as an encoding
effect.

\paragraph{A screen we require elsewhere had never been run on this table.}
the main paper's instrument section treats the echo screen as a validity requirement, and the main paper's limitations section says the same. Applied here for the first time, it rejects
four of the nine encoded cells: sft/fullwidth, dpo/zero-width, rlvr/fullwidth and
rlvr/zero-width, with displacements of $0.11$ to $0.26$ against half-widths of
$0.09$ to $0.13$. Homoglyph clears at all three stages by an order of magnitude
($0.003$ to $0.011$), which is why the table above is built on it. The reason the
screen had not been run is worth stating rather than hiding: it compares the
reported gap against the gap over non-echoing cells, so it needs \emph{both}
arms' per-item verdicts, and this experiment predates the benign arm being
written to disk at all. A screen adopted in one section does not reach backwards
into a table written before it existed, and nothing in our pipeline noticed. We
count this as defect~(10) of the main paper's instrument section and it is the second we
found by turning one of our own instruments on one of our own results.

\paragraph{The judge is the instrument.} Every number here is an LLM judge's
verdict. We validate it three ways and it remains, at the end, a model judging a
model. It separates plaintext harmful from plaintext benign
(the main paper's instrument section); it has a negative control whose ground truth holds by
construction (item 4); and we checked its verdicts against two independent
annotators drawn from model families other than the judge's own, on $200$ items
sampled blind and stratified on the judge's verdict so that both error directions
are estimable. The annotators were given the judge's own definition, under which
a response that merely echoes or is irrelevant to the request counts as a
refusal, so that a disagreement about wording could not be recorded as judge
error. Agreement is $0.90$--$1.00$ per stratum; the judge calls a refusal on
something an annotator calls engagement in $0.040$ and $0.050$ of cases, and
misses a refusal in $0.051$ and $0.043$. The two annotators agree with each other
on $0.930$ of the items both labelled, and nine of their twelve disagreements lie
on the refusal-versus-echo boundary, which sits inside the judge's ``refused''
and cannot move any rate we report. \textbf{This is model judgement and not human
annotation, and we label it as such.} One family declined to label $27$ of the
$200$ items, concentrated on the harmful arm ($0.20$ against $0.08$ benign) and
on the cells the judge called refusals; those items are excluded and the
concentration is reported, because a completion rate alone would have hidden it.
Earlier work with this judge model, on a different refusal rubric and a different
response distribution, reached $\kappa = 0.79$ against a blind human annotator;
we cite that as inherited context and not as validation of the present
configuration, since the rubrics differ precisely on the echo case.

\paragraph{Nothing here is held out.} $n{=}100$ is the whole of the
JailbreakBench harmful set and the whole of its theme-matched benign counterpart.
Every threshold we set (the decode-similarity and overlap cuts, the probe
licensing floor, the control-floor width, the read percentile) and every rung,
layer and position we selected was set or selected on these same $200$ items. No
number in this paper has been evaluated on data that took no part in producing
it. Defect (9) is the sharp instance of this and it is not the only one: what
that probe did with items, the pipeline as a whole does with knobs. The
generalisation this most threatens is the rung selection, since a screen tuned
and applied on one corpus can select the rung that suits it. Three further
corpora sit unused in our own data directory, and a replication that sets the
knobs on one and reports on another is the next run rather than a future
direction.

\section{Further limitations}
\label{sup:limits}

\subsection{Four further limitations}

\paragraph{$n{=}100$ per cell, and runs that do not repeat exactly.}
Differences below roughly $0.15$ are not separable, which is why we report a
noise null for every cross-model spread and a confidence interval for every gap
change. The measured run-to-run spread reported in the main paper's method section is
$0.01$--$0.06$, inside that bound, so no figure here rests on a difference the
pipeline cannot reproduce. It is smaller than the sampling bound only because
the verdict flips cancel, which is a property of these conditions and not a
guarantee about others.

\paragraph{Equalisation versus saturation is unresolved.} Encoded harmful
refusal sits at $0.91$--$0.99$ on all four models, so two readings fit every
cell we report: the protocol \emph{equalises} the two arms, or it
\emph{saturates} the harmful arm at a ceiling the benign arm climbs toward.
\S\ref{sup:defects} states that tension in full. What we add here is what it
would take to settle it, because $n{=}100$ against a ceiling cannot: either a
model whose encoded harmful refusal sits well below $1.00$, or a graded refusal
measure in place of our binary one. We have neither. Our claim is deliberately
the weaker one that holds under both readings.

\paragraph{Whether the harm information is lost or merely unused is open, and
our attempt to settle it failed.} The natural next question is whether the
protocol destroys the model's representation of harm or leaves it intact and
unread. Our attempt to answer it failed, and we report the failure as defect (9) of
the main paper's instrument section rather than as a result. What survives it is a bound
rather than an answer: with items held out, the harm direction transfers
\emph{below} each model's own within-plaintext baseline, which points toward a
representation degraded by the protocol rather than an intact one that refusal
ignores. We do not claim that either. It rests on $100$ items per arm, and the
two readings it must separate, degraded and intact-but-unread, call for
opposite remedies, so distinguishing them deserves an experiment designed for
it. That experiment needs more items than we have and a causal arm we have not
run.

\paragraph{An error we made, kept in.} We concluded that the SFT checkpoint
``behaves like'' one of the other models, on the strength of their matching
encoded arms. With the plaintext arm they are different phenomena: the
encoding-induced benign excess is $+0.89$ for one and $+0.34$ for the other, and
the SFT checkpoint already refuses $0.45$ of benign \emph{plaintext}. The claim
was made from the encoded arm alone, which is the exact inference this paper
argues is unsound. We record it because it shows the failure is easy to make with the data
in hand and a conclusion in mind.

\subsection{The internals attempt we withdrew}

\paragraph{Reading refusal inside the model.} \citet{arditi2024refusal} show
refusal is mediated by a single direction that is necessary and sufficient
across models, and select that direction by intervention (ablating it and
checking that refusal is bypassed) rather than by correlational fit. We adopt
their fitting method for the content probe we report on in
the main paper's instrument section, and we take their selection criterion as the standard
our probe does not yet meet: our licensing is correlational, which is why we
screen against a decode-impossible floor and report the probe's reach as two
encodings rather than the fourteen a permutation test licenses. That gap is the
main reason this paper reports behaviour rather than mechanism.

\citet{zhao2025llmsencode} find that harmfulness and refusal are represented
separately, harmfulness at the final instruction token and refusal downstream of
it. That separation is what turns our behavioural result into a question rather
than an answer. A protocol could leave the upstream quantity intact and disrupt
only the downstream use of it, or it could degrade the representation itself,
and the two call for opposite remedies: a read-out repair in the first case,
more or better safety data in the second. Refusal rates cannot tell them apart,
and our own attempt to do so with a transferred probe is defect (9) of
the main paper's instrument section rather than a finding. We think it is the most valuable
open question this paper leaves.

\end{document}